\documentclass[aps,prb,superscriptaddress,twocolumn,showpacs,natbib]{revtex4}
\usepackage{color}
\usepackage{graphicx}
\usepackage{cancel}
\usepackage{bm}
\usepackage{amsmath}
\usepackage{amssymb}
\def  \bsig    {\mbox{\boldmath$\sigma$}}

\def    \bta          {\mbox{\boldmath$\tau$}}

\begin{document}

\title{Optical spin injection in graphene with Rashba spin-orbit interaction}

\author{M. Inglot}
\affiliation{Department of Physics, Rzesz\'ow University of Technology,
al.~Powsta\'nc\'ow Warszawy 6, 35-959 Rzesz\'ow, Poland}

\author{V. K. Dugaev}
\affiliation{Department of Physics, Rzesz\'ow University of Technology,
al.~Powsta\'nc\'ow Warszawy 6, 35-959 Rzesz\'ow, Poland}
\affiliation{Departamento de F\'isica and CFIF, Instituto Superior T\'ecnico,
Universidade de Lisboa, av.~Rovisco Pais, 1049-001 Lisbon, Portugal}

\author{E. Ya. Sherman}
\affiliation{Department of Physical Chemistry, Universidad del Pa\'is Vasco UPV-EHU, 48080 Bilbao, Spain}
\affiliation{IKERBASQUE Basque Foundation for Science, Bilbao, Spain}

\author{J. Barna\'s}
\affiliation{Faculty of Physics, Adam Mickiewicz University, ul. Umultowska 85,
61-614 Pozna\'n, Poland}
\affiliation{Institute of Molecular Physics, Polish Academy of Sciences, ul. M. Smoluchowskiego 17,
60-179 Pozna\'n, Poland}

\begin{abstract}
We calculate the efficiency of infrared optical spin injection in single-layer graphene
with Rashba spin-orbit coupling and for in-plane magnetic field. The injection rate in the photon frequency range
corresponding to the Rashba splitting is
shown to be proportional to the ratio of the Zeeman and Rashba splittings. As a result,  large spin polarization can be controllably
achieved for experimentally available values of the spin-orbit coupling and in magnetic fields below 10 Tesla.
\end{abstract}

\date{\today}

\maketitle

\section{introduction}

Graphene -- a two-dimensional hexagonal lattice of carbon atoms -- was discovered about eight
years ago \cite{novoselov05,zhang05,katsnelson}
and is now one of the most promising materials for future nanoelectronics. The high application potential
of this novel material  is associated with
some peculiarities of its electronic and phonon transport properties \cite{balandin} as well as with
its outstanding mechanical \cite{ranjbartoreh11,chen08} and optical properties \cite{Falkovsky}.
Optoelectronic properties of graphene are also very promising for applications \cite{Ferrari}.
Moreover, owing to a very long spin relaxation time, which is
expected due to a very weak spin-orbit interaction, graphene is also attractive for applications in
spin electronics (see, e.g., Ref.~[\onlinecite{guimaraes10}]).

However, to utilize the outstanding
properties of graphene for spin-dependent transport, one needs to have a reliable method of controllable
spin injection and spin manipulation. The possibility
of a relatively strong Rashba spin-orbit coupling
has been reported in Refs.~[\onlinecite{dedkov08,varykhalov08}] for graphene deposited on a Ni (or Ni/Au) substrate.
Such a strong spin-orbit coupling formally enables spin manipulation in graphene. However,
even in the absence of a substrate leading to strong spin-orbit coupling,
experiments report spin relaxation time on the timescale of the order of or less than one
nanosecond \cite{Wees,Kawakami,Guntherodt}, which makes
applications of graphene in spin electronics rather difficult. In all the experiments aimed at the measurements of
spin relaxation time, spins are injected {\it en masse} from a
ferromagnetic contact giving rise to some charge/spin density distribution, which influences its
subsequent dynamics. Several theoretical approaches (see for example Refs.~[\onlinecite{dugaev11,MWWu,Fabian,Fratini,Ochoa}])
have been proposed to describe spin relaxation. However, most of them demonstrated spin relaxation time much
longer than that observed experimentally.

There are several experimental techniques which can be used to manipulate and control electron spin in graphene.
For example, spin
current and spin density in graphene nanodisks can be manipulated by varying length of the
corresponding zigzag edge~\cite{ezawa10}. Quantum pumping of Dirac fermions and spin current in a
monolayer graphene in perpendicular
magnetic field, with the gate voltage as a control parameter, has been
proposed in Ref.~[\onlinecite{tiwari10}]. Furthermore, the method of spin current
generation in a monolayer graphene through adiabatic quantum pumping by two
oscillating in time potentials has been described in Refs.~[\onlinecite{zhang11,greenbaum07}]
and for the bilayer graphene in Ref.~[\onlinecite{Feng11}].

It is well-known that spin-orbit  coupling can lead to a direct
optical spin injection - the technique extensively used in the physics of
semiconductors \cite{SObook}. For graphene, the spin-orbit coupling influences the optical
response in the infrared frequency range \cite{Zulicke}.
In this paper
we consider the infrared optical spin injection  in a single-layer graphene by linearly polarized light.
The graphene is assumed to be deposited on a substrate which leads to the Rashba
spin-orbit coupling.
Due to this interaction, the electronic spectrum of graphene
near the Dirac points splits into four bands with parabolic
dependence on the electron momentum at small wave vectors and linear dispersion at large wave vectors\cite{Rashba09}.
Splitting of the subbands is determined by the spin-orbit coupling strength.
We show that optical spin injection becomes allowed in the presence of an external magnetic field,
and the injection efficiency is of the
order of the ratio of the Zeeman splitting and the spin-orbit coupling matrix
element. By modifying the infrared light frequency, one can change the
absorption region in the momentum space, and thus control the spin
injection.

In Sec.~2 we derive some general formula for optical spin injection efficiency in graphene.
Numerical results on spin injection rate are presented and described in Sec.~3.
Summary and final conclusions are in Sec.~4.

\section{Spin injection rate and efficiency}

We assume an external magnetic field $\mathbf{B}$ oriented in the graphene plane.
Hamiltonian describing the low energy electron excitations near the Dirac point $K$
in graphene with the Rashba spin-orbit interaction takes then the form~\cite{kane05}
\begin{eqnarray}  \label{1}
\hat{H} =v(\bta \cdot \mathbf{k})+g(\mathbf{B}\cdot \bsig) +
\lambda (\tau_x\sigma _y-\tau _y\sigma _x),
\end{eqnarray}
where $g=g_L\mu _B/2$, $\lambda =\alpha/2$ with $\alpha$ being the coupling constant
of Rashba spin-orbit interaction,\cite{Rashba09} and $g_L$ is the Land\'e factor. The
matrices $\bta$ and $\bsig$ are the Pauli matrices in the sublattice
and spin space, respectively. The third term of the above Hamiltonian
stands for the Rashba spin-orbit coupling induced by the substrate. Note, the first term is
diagonal in the spin space and for abbreviation the corresponding  unit matrix is not written explicitly.
Similarly, the
second term is diagonal in the sublattice space and the corresponding  unit matrix is not written explicitly, too.

Electronic spectrum corresponding to the  Hamiltonian (\ref{1}) consists of four energy bands.
In the limit of weak magnetic field, $gB/\lambda \to 0$, this
spectrum is described by the formulas
\begin{equation}
E_{n{\bf k}}^{(0)}= \pm\lambda \pm(\lambda^2+v^2k^2)^{1/2},
\end{equation}
with all possible combination of the $+$ and $-$ signs, and the index $n$ labeling the bands
in the order of increasing energy (see Fig.~1).

As in  the usual two-dimensional electron gas with Rashba spin-orbit interaction,
the expectation value of the spin $z$-component in eigenstates of Hamiltonian (1) for $B=0$ is equal to zero.
However, unlike to the two-dimensional electron gas, the expectation value of the in-plane spin
depends on the wave vector and is relatively small for low-energy electron states.
Indeed, the expectation value,
$\langle\Psi^{(0)}_{n\mathbf{k}}|\bsig|\Psi^{(0)}_{n\mathbf{k}}\rangle$,
of electron spin in the state
$\Psi^{(0)}_{n\mathbf{k}}$ for $B=0$
is given by \cite{Rashba09}
\begin{equation}
{\bf s}\equiv \langle\Psi^{(0)}_{n\mathbf{k}}|\bsig|\Psi^{(0)}_{n\mathbf{k}}\rangle
=\frac{\xi v \left({\bf k}\times\hat{\bf z}\right)}
{\sqrt{\lambda^{2}+v^{2}k^{2}}},
\label{sigmak}
\end{equation}
where $\xi=\pm1$ is the band index, $\xi=1$ for $n=2,3$ and $\xi=-1$ for $n=1,4$
(cf. Ref.~[\onlinecite{Rashba09}]).
Thus, when $vk\ll \lambda $, the expectation value of spin is small, $|{\bf s}|\ll 1$.
Moreover, the spins are perpendicular to the wave vectors, similarly as in two-dimensional electron gas.
We note that the upper index (0) at the eigenfunctions and eigenenergies indicates they are for $B=0$.

\begin{figure}[ht]
\includegraphics[scale=0.45]{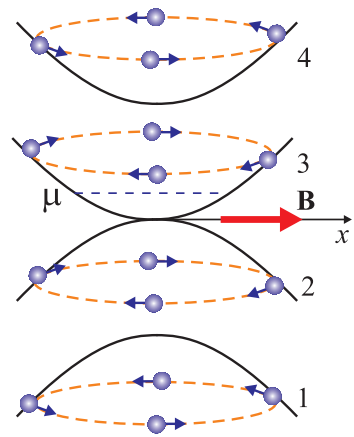}
\caption{(Color online). Band structure of graphene with a
Rashba spin-orbit interaction in the $gB/\lambda\ll 1$ limit.
Arrows show the ${\mathbf k}$-dependent orientation of electron spins in the eigenstates
of the Hamiltonian. Here we consider only the vicinity of the $K$ Dirac point, taking into account that
the other valley gives exactly the same result for light absorption and spin injection.}
\label{fig1}
\end{figure}

In the following, we take the in-plane magnetic field $\mathbf{B}$  along the
$x$-axis and assume it is rather weak, $gB/\lambda \ll 1$. The former assumptions
justifies the absence of Landau quantization, while the latter condition assures that the
band dispersion is only weakly perturbed by the static magnetic field, and the resulting
spin injection is linear in the applied magnetic field ${\bf B}$.
The four-band structure is presented in Fig.~\ref{fig1}, which also shows
the assumed position of the Fermi level $\mu $.

Taking into account the first term of Eq.~(1), Hamiltonian describing interaction of graphene with an external
periodic electromagnetic field
$\mathbf{A}(t)=\mathbf{A}_0e^{-i\omega t}$ can be written as
\begin{eqnarray}  \label{2}
\hat{H}_A=-\frac{ev}{\hbar c}\, (\bta\cdot\mathbf{A}).
\end{eqnarray}
This periodic perturbation leads to electron transitions between
the bands shown in Fig.~1. The spin states of electrons involved in the transitions are then modified accordingly.
Without loss of generality, we assume in the following that the electromagnetic field is oriented along the $y$-axis,
$\mathbf{A}_0=(0, A_0, 0)$.

The total absorption rate of photons can be calculated as the sum of all allowed transitions,
\begin{eqnarray}  \label{3}
I(\omega )=\sum _{nn^{\prime}}I^{n\to n^{\prime}}(\omega ),
\end{eqnarray}
where $I^{n\to n^{\prime}}(\omega)$ corresponds to the absorption associated with the
transitions of electrons from the subband $n$ to the subband $n^{\prime}$, which can be calculated
from the Fermi  golden rule as
\begin{eqnarray}
\label{4}
I ^{n\to n^{\prime}}(\omega ) =\frac{2\pi }{\hbar }\int \frac{d^2\mathbf{k}}
{(2\pi)^2}\;
\left|\langle\Psi_{n\mathbf{k}}|\hat{H}_{A}|\Psi_{n^{\prime}\mathbf{k}}\rangle\right|^2  \nonumber \\
\times \delta (E_{n\mathbf{k}}+\hbar \omega -E_{n^{\prime}\mathbf{k}})\,
f(E_{n\mathbf{k}})\, [1-f(E_{n^{\prime}\mathbf{k}})].
\end{eqnarray}
Here, $\Psi_{n\mathbf{k}}$ and $E_{n\mathbf{k}}$ are the
eigenfunctions and eigenvalues of the total Hamiltonian \eqref{1}, and $f(E_{n\mathbf{k}})$
is the corresponding Fermi distribution function.
% In the $gB/\lambda \ll 1$ limit,
%the effect of static magnetic field $B$ on the light absorption is only marginal
%and can be neglected.

It is convenient to introduce an independent of the system parameter $I_{0}$, defined as
\begin{equation}
I_{0}=\frac{\omega}{4}\left(\frac{e}{\hbar c}\right)^{2}A_{0}^{2},
\label{I0i}
\end{equation}
and rewrite  Eq.~\eqref{3} as
\begin{equation}
I(\omega )=I_{0}\,\sum _{nn^{\prime}}\widetilde{I}^{n\to n^{\prime}}(\omega)\equiv I_{0}\widetilde{I}(\omega),
\end{equation}
with the system-dependent functions $\widetilde{I}^{n\to n^{\prime}}(\omega)$.
Since  $A_{0}^{2}$ in Eq.~\eqref{I0i} is related to the incident flux $q$
of $y$-polarized photons by the formula $A_{0}^{2}=4\pi \hbar cq/\omega$,
Eq.~\eqref{I0i} can be presented in the form
\begin{equation}
I_{0}=\frac{\pi e^{2}}{\hbar c}\, q.
\end{equation}
The ratio $I_{0}/q=\pi e^{2}/\hbar c$ corresponds to the absorption coefficient of
graphene without Rashba spin-orbit coupling. \cite{katsnelson,gusynin06,kuzmenko08}
In the limit of large frequency, $\hbar\omega\gg\lambda $,
the absorption rate (8) is constant and does not depend on frequency,  like in the case of
graphene with zero Rashba coupling, $I(\omega )\to I_{0}$. Thus, $\widetilde{I}(\omega)$ can be
considered as a ratio of absorption coefficients for graphene
with Rashba spin-orbit interaction and of graphene without Rashba interaction. In other words,
 $\widetilde{I}(\omega)$  is the absorption coefficient normalized to that for graphene without Rashba interaction.

Now, let us define the spin injection rate for the $i$-th component of the spin density. Following
Eq.~\eqref{4}, we write
\begin{eqnarray}
J_{i}^{n\to n^{\prime }}(\omega ) &=&\frac{2\pi }{\hbar }
\int \frac{d^{2}\mathbf{k}}{(2\pi )^{2}}\;
\left|\langle\Psi_{n\mathbf{k}}|\hat{H}_{A}|\Psi_{n^{\prime}\mathbf{k}}\rangle\right|^2 \nonumber
\label{5} \\
&&\hspace{-2cm}
\times \left(
 \langle \Psi_{n^{\prime}\mathbf{k}}|\sigma _{i}|\Psi_{n^{\prime}\mathbf{k}} \rangle
-\langle \Psi_{n\mathbf{k}}|\sigma _{i}|\Psi_{n\mathbf{k}} \rangle
\right)  \nonumber \\
&&\hspace{-2cm}
\times\delta (E_{n\mathbf{k}}+\hbar\omega - E_{n^{\prime}\mathbf{k}})
f(E_{n\mathbf{k}})\,[1-f(E_{n^{\prime }\mathbf{k}})].
\end{eqnarray}
Similarly to the case of absorption, we introduce the total spin injection rate as
$J_{i}(\omega )=\sum_{n,n^\prime}J_{i}^{n\to n^{\prime }}(\omega )$ and write
$J_{i}(\omega )=I_0\widetilde{J}_i(\omega)$ and
$J_{i}^{n\to n^{\prime }}(\omega )=I_0\widetilde{J}_{i}^{n\to n^{\prime }}(\omega )$.
Thus, $\widetilde{J}_i(\omega)$ and $\widetilde{J}_{i}^{n\to n^{\prime }}(\omega )$ can be considered as
normalized to $I_0$ spin injection rates.
Before discussing numerical results based on the above formula, let us discuss briefly physical origin of the spin injection.

In the absence of magnetic field, symmetry of the matrix elements and
spin expectation values (as shown in Fig.~\ref{fig1}) as well as
the independence of energy $E_{n\mathbf{k}}$ on the momentum orientation
lead to zero spin injection rate, as required by the time-reversal symmetry.
In an in-plane magnetic field, in turn, each subband is shifted in energy,
$E_{n{\mathbf k}}-E^{(0)}_{n{\mathbf k}}=
g\left(\langle\Psi^{(0)}_{n\mathbf{k}}|{\bm\sigma}|\Psi^{(0)}_{n\mathbf{k}}\rangle\cdot{\mathbf B}\right).$
As a result, the lines of energy conservation, $E_{n\mathbf{k}}+\hbar \omega -E_{n^{\prime}\mathbf{k}}=0,$
for the transitions changing the electron spin, such as $1\rightarrow 3$ and $2\rightarrow 4$,
are not simple circles anymore and
acquire a distortion of the order of $(gB/\lambda)\cos\varphi$,
where $\varphi$ is the angle between $\mathbf{k}$ and the $x$-axis.
In addition, the 4-component wave functions are modified in the first order perturbation as
\begin{equation}
\Psi_{n{\mathbf k}}-\Psi^{(0)}_{n{\mathbf k}}=
{g}\sum_{n^{\prime }}
\frac{\left< \Psi^{(0)}_{n^{\prime}{\mathbf k}}\right|
\left({\bm\sigma}\cdot{\mathbf B}\right)
\left|\Psi^{(0)}_{n{\mathbf k}}\right>}
{E^{(0)}_{n\mathbf k}-E^{(0)}_{n^{\prime}{\mathbf k}}}\Psi^{(0)}_{n^{\prime}}.
\end{equation}
Accordingly, expectation values of the spin components
$\langle\Psi_{n{\mathbf k}}|\sigma _{i}|\Psi_{n{\mathbf k}}\rangle $ and of the interband
matrix elements  $\langle\Psi_{n{\mathbf k}}|\hat{H}_{A}|\Psi_{n^{\prime}{\mathbf k}}\rangle$
acquire first-order modification in the applied magnetic field,
which results in a nonzero spin injection.

\section{Infrared absorption and spin injection: numerical results}

Now we present some numerical results on the absorption of linearly polarized light and the associated spin injection.
In our calculations we assumed the temperature $T=1$~K.

\begin{figure}[ht!]
\includegraphics[scale=0.45]{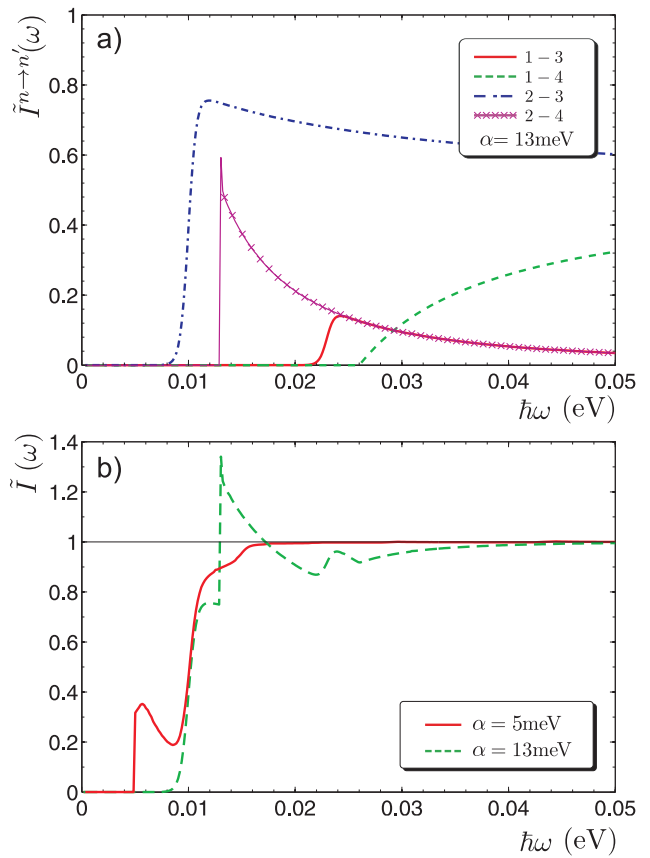}
\caption{(Color online). (a) The normalized absorption coefficients corresponding to indicated
interband transitions, calculated  for $\alpha=2\lambda=13$ meV.
(b) The total normalized absorption coefficient for
two different values of the  Rashba spin-orbit parameter, as indicated.
Both figures are calculated for magnetic field $B=5$ T
and for the chemical potential $\mu=5$ meV.}
\label{fig2}
\end{figure}

Let us begin with the normalized absorption coefficients presented in Fig.~\ref{fig2}
as a function of the frequency $\omega$.
Figure~\ref{fig2}a shows the normalized absorption coefficients for individual inter-band transitions, while
Fig.~\ref{fig2}b shows the total normalized absorption coefficient for two different values of the
Rashba parameter $\alpha$.  In the latter case, the thin solid line corresponds
to the absorption coefficient in the absence of Rashba coupling.

As one can see in Fig.~\ref{fig2},
the spin-orbit coupling strongly modifies the absorption, in agreement with the results
of Ref.~[\onlinecite{Zulicke}]. The frequency threshold for the interband transitions
is determined by the Pauli blocking and also
depends on the chemical potential $\mu$ and the spin-orbit splitting $2\lambda$.
In the case considered here, $\mu<2\lambda$, the transitions of highest frequency
occur between the $n=1$ and $n^{\prime}=4$ bands and start at the $K$-point with zero  matrix
element. At high frequencies, $\hbar\omega\gg\lambda$,
the total absorption coefficient approaches that for
a pure single-layer graphene without spin-orbit coupling.

\begin{figure}[ht!]
\includegraphics[scale=0.45]{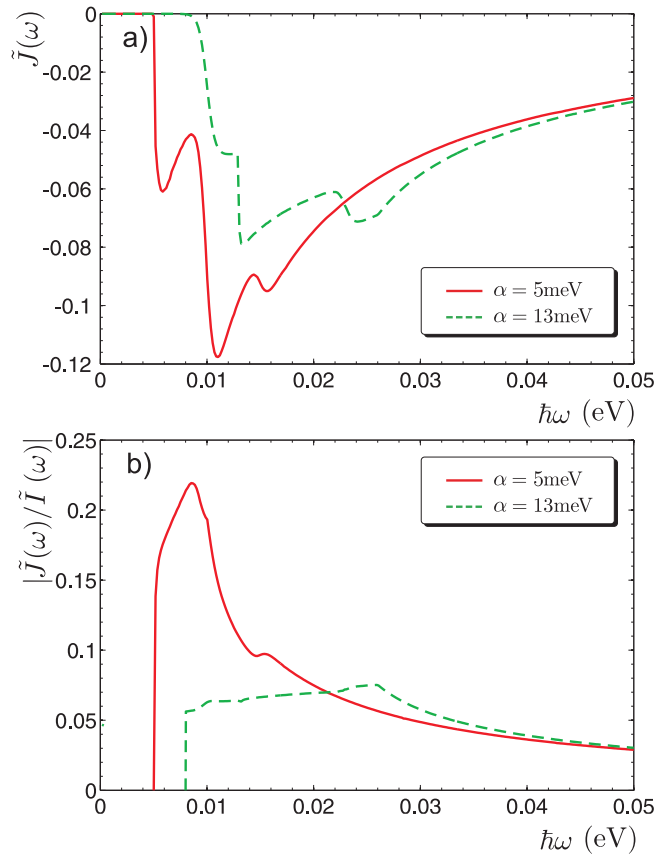}
\caption{(Color online). (a) The total injection rate for the spin component $\sigma_x$
in a magnetic field of $B=5$ T parallel to the $x$-axis and external electromagnetic field
polarized along the $y$-axis. The dashed green (solid red) line corresponds to the coupling
constant $\alpha=2\lambda=13$ meV ($\alpha=2\lambda=5$ meV).
All results are for the chemical potential $\mu=5$ meV.
(b) Spin injection efficiency for spin change per absorbed photon for the same parameters and injection geometry.}
\label{fig3}
\end{figure}

Let us consider now numerical results on spin injection shown in Fig.~\ref{fig3} for
the magnetic field $B=5$ T, which corresponds
to the Zeeman splitting $2gB$ of approximately 0.6 meV. We show there only the injection rate
for the spin $x$-component,
and for clarity we simplified there  the notation by omitting the index $i$, $\tilde{J}_x(\omega)\equiv \tilde{J}(\omega)$.
Figure~\ref{fig3}a shows the total normalized spin injection rate as a function of frequency for two different values of the
Rashba coupling parameter.
The dependence of the spin injection rate
on the light frequency is rather complicated due to several interband transitions involved and
complex dependence of the matrix elements on the transition frequency.
For the spin-conserving transitions between the states
characterized by the same $\xi$ in Eq.~\eqref{sigmak}, such as $1\to4$ and $2\to3$, the main contribution
to the spin injection comes from changes in the expectation values of
$\langle\Psi_{n\mathbf{k}}|\bsig|\Psi_{n\mathbf{k}}\rangle$, while for the other transitions
all the changes in the system make comparable contributions. Note, the spin injected is opposite to the magnetic field.

The spin injection efficiency can be defined as the average spin injected by a single photon.
This efficiency is given by $|\widetilde{J}\omega)/\widetilde{I}(\omega)|$, and is shown in Fig.~\ref{fig3}b for the same
Rashba parameters as in Fig.~\ref{fig3}a.
As one can see in Fig.~\ref{fig3}b, the efficiency can reach 0.2 per incident photon. In general,
the injection rate is of the order of $gB/\lambda$, and can be manipulated by changing the photon frequency
in the range of the order of $\lambda$.
Since the transitions are rather complicated, the ratio $gB/\lambda$ should be considered as an
order-of-magnitude estimate only.

Similar spin injection, though considerably weaker, can be obtained
for the electric field along the magnetic field, i.e. along the $x$-axis.

\section{Summary and conclusions}

We have considered theoretically spin injection in single-layer graphene in the presence of Rashba
spin-orbit coupling and in-plane external magnetic field. We have found that the spin
injection is efficient at frequencies of the order of spin-orbit band
splitting, with the efficiency  being of the order of the ratio of Zeeman and Rashba splittings.

For experimentally achievable parameters of the spin-orbit coupling and magnetic
field, the injection efficiency can achieve 0.2 per absorbed photon.
This result shows that optical spin injection opens a way for a controllable and
efficient method of spin density and spin current generation
in graphene, not requiring presence of any ferromagnetic contact.

\section*{Acknowledgements}

This work is partly supported by the National Science Center in Poland as a
research project in years 2011 -- 2014 and Grants Nos.~DEC-2011/01/N/ST3/00394
and DEC-2012/06/M/ST3/00042.
The work of EYS
was supported by the University of Basque Country UPV/EHU under
program UFI 11/55, Spanish MEC (FIS2012-36673-C03-01), and ''Grupos
Consolidados UPV/EHU del Gobierno Vasco'' (IT-472-10).

\end{document}